\documentclass[12pt,a4paper]{article}
\usepackage{graphicx}
\usepackage[latin2]{inputenc}
\usepackage{latexsym}
\usepackage{epsf}
\usepackage{epsfig}
\usepackage{cite}

\def\bea{\begin{eqnarray}}  
\def\eea{\end{eqnarray}}  
\def\bc{\begin{center}}
\def\ec{\end{center}}

\def\simlt{\stackrel{<}{{}_\sim}}

\begin{document}
\pagestyle{empty}
\begin{flushright}
IFT-23/2005\\ 
DESY-05-193\\
CERN-PH-TH/2005-179\\ 
{\bf \today}
\end{flushright}
\vspace*{5mm}
\begin{center}

{\large {\bf Double protection of the Higgs potential}}\\
\vspace*{1cm}

{\bf Zurab Berezhiani}$^{\rm a)}$\footnote{Email:zurab.berezhiani@cern.ch}, 
{\bf Piotr~H.~Chankowski}$^{\rm b)}$\footnote{Email:chank@fuw.edu.pl}, 
{\bf Adam Falkowski}$^{\rm b),c)}$\footnote{Email:afalkows@fuw.edu.pl},\\
and 
{\bf Stefan~Pokorski}$^{\rm b),d)}$\footnote{Email:pokorski@fuw.edu.pl}
\vspace{0.5cm}
 
a) Dipartimento di Fisica, Universita di L'Aquila, 67010 Coppito AQ and\\
   Laboratori Nazionali del Gran Sasso, 67010 Assergi AQ, Italy\\
b) Institute of Theoretical Physics, Warsaw University, Ho\.za 69, 00-681,
   Warsaw, Poland\\
c) Deutsches Electronen-Synchrotron DESY, Notkestrasse 85, 22607 Hamburg,
   Germany\\
d) CERN Theory Division, CERN, Geneva 23, Switzerland\\

\vspace*{1.7cm}
{\bf Abstract}
\end{center}
\vspace*{5mm}
\noindent
{ 
A mechanism of double protection of the Higgs potential, by supersymmetry and 
by a global symmetry, is investigated in a class of supersymmetric models with 
the $SU(3)_C\times SU(3)_W \times U(1)_X$ gauge symmetry. In such models the 
electroweak symmetry can be broken with no fine-tuning at all. 
}
\vspace*{1.0cm}
\date{\today}


\vspace*{0.2cm}
 
\vfill\eject
\newpage

\setcounter{page}{1}
\pagestyle{plain}

Stabilizing the Higgs potential is the central motivation for most 
extensions of the standard model. In softly broken supersymmetry the 
quadratic sensitivity of the Fermi scale to the ultraviolet (UV) cutoff  
is removed to all orders in perturbation theory. In generic supersymmetric  
models the Higgs potential mass parameter depends quadratically on the 
soft supersymmetry breaking scale $M_{\rm soft}$ and logarithmically on the 
cut-off $\Lambda_{UV}$. For example, in the MSSM the one-loop corrections 
which lead to the electroweak symmetry breaking are dominated by the top 
sector contribution and one approximately has:
\begin{eqnarray}
m_H^2\approx m_0^2-{3\over8\pi^2}y_t^2M_{\rm soft}^2
\ln{\Lambda_{UV}^2\over M_{\rm soft}^2}~,\label{eqn:mssmft}
\end{eqnarray}
where $y_t$ is the top quark 
Yukawa coupling. This mechanism of radiative electroweak symmetry breaking 
strongly links the electroweak scale with $M_{\rm soft}$. Indeed, the 
tree-level term $m_0^2$ contains a supersymmetry breaking contribution of 
order $M_{\rm soft}^2$ (and the $\mu$-term contribution, which should be of 
the same order as $M_{\rm soft}$), and for $\Lambda_{UV}$ close to the GUT 
scale the one loop term is also of that order. However, in view of the 
existing experimental constraints, such relation appears to be unsatisfactory. 
The lower limit on the Higgs boson mass and the precision electroweak data 
put a lower bound on $M_{\rm soft}$ of order of 1~TeV. In consequence, since 
$m_H^2\approx -M_Z^2/2$, the cancellation between the tree-level and one-loop 
terms in the Higgs potential must be at least 1 part to 100 - the fact known 
as the ``supersymmetric fine-tuning problem''.

The following two features would be welcome to improve this picture. The 
underlying physics should forbid the tree-level Higgs mass parameter. This 
would also have an advantage of avoiding the $\mu$-problem of the MSSM.
Secondly, $\Lambda_{UV}$ in (\ref{eqn:mssmft}) should be replaced with 
another scale of order $M_{\rm soft}$. This would lead to 
$m_H^2\sim M_{\rm soft}^2/16\pi^2$ and the correct value of $m_H^2$ for 
$M_{\rm soft}$ of order 1~TeV.  

The supersymmetric fine-tuning problem has stimulated several authors to 
look for alternatives to supersymmetry. The little Higgs models \cite{LITHIG} 
revive the idea of the Higgs doublet being a pseudo-Goldstone boson of 
some global symmetry spontaneously broken at a scale ${\cal O}(1$~TeV). 
However, the scale of the global symmetry breaking is usually linked to the 
mass scale of new gauge bosons, $W^{\pm\prime}$ and/or a $Z^\prime$, and is 
constrained by 
precision  electroweak data. In consequence, the fine-tuning in the Higgs 
potential is at least as large as in supersymmetric models \cite{CASAS,STSCH}. 

In this paper we explore the idea of double protection of the Higgs potential.
This mechanism operates in supersymmetric models in which the Higgs doublet 
is a pseudo-Goldstone boson of an approximate global symmetry spontaneously 
broken at ${\cal O}(1$~TeV). One role of the additional symmetry is to forbid 
the tree-level Higgs mass term. In addition, its interplay with supersymmetry
removes also the logarithmic dependence of $m_H^2$ on $\Lambda_{UV}$ at one 
loop: $\Lambda_{UV}$ gets replaced there with the mass scale of additional 
vector-like quark multiplets which is of the order of the spontaneous global 
symmetry breaking scale $f$. As a result, the dominant one-loop contribution 
to the Higgs potential mass parameter is finite and takes the form:
\begin{eqnarray}
m_H^2\approx-{3\over8\pi^2}y_t^2\left[(M_{\rm soft}^2+f^2)
\ln(M_{\rm soft}^2+f^2) - M_{\rm soft}^2\ln M_{\rm soft}^2
- f^2\ln f^2\right]~.
\end{eqnarray}
This vanishes in the limit of unbroken supersymmetry 
$M_{\rm soft}\rightarrow0$ as well as for unbroken global symmetry 
$f\rightarrow0$. 

The idea of the double protection has been explored in  
ref.~\cite{CHFAPOWA} in a model proposed in ref.~\cite{BICHGA}. However 
the unattractive feature of this model 
is that the scale of the $SU(3)$ symmetry
breaking is linked to the mass of the $Z^\prime$ boson. The allowed 
parameter space is then very limited and the fine-tuning remains as large as 
in the MSSM, although for different reasons. In this paper we discuss the 
mechanism of double protection in a class of models in which the global 
$SU(3)$ is a natural consequence of a $SU(3)$ gauge symmetry. Furthermore, 
the scale $f$ of spontaneous breaking of the global symmetry is not related 
to the scale $F$ of spontaneous gauge symmetry breaking and the experimental 
limits on the masses of new gauge bosons do not constrain $f$. For  
$f\simlt1~{\rm TeV}\ll F$, the electroweak symmetry can be broken with no 
fine-tuning at all. 

We consider a class of supersymmetric models with 
$SU(3)_C\times SU(3)_W\times U(1)_X$ gauge symmetry. The electroweak 
$SU(2)_W \times U(1)_Y$ group is a subgroup of $SU(3)_W\times U(1)_X$ and the 
matter and Higgs fields are extended to $SU(3)_W$ multiplets. Several models 
of this kind exist in the literature \cite{SCHM,FRAMP} and several others 
can be constructed. They differ in the assignment of particles to 
$SU(3)_W\times U(1)_X$ representations as well as in existence of additional 
exotic matter multiplets. The assignment can be such that all anomalies 
cancel \cite{SCHM,FRAMP}. In this letter we concentrate only on the most 
universal features of such models.

We shall require that the Higgs sector has global symmetry 
$SU(3)_1\times SU(3)_2$ whose diagonal subgroup is the gauge $SU(3)_W$ group. 
This can be easily achieved if the Higgs sector consists of two pairs of 
Higgs multiplets:
\begin{eqnarray}
\Phi_D~, \phantom{aa} \Phi_U~,\phantom{aaaa}{\rm and} \phantom{aaaa}   
{\cal H}_d~, \phantom{aa} {\cal H}_u~,
\end{eqnarray}
where $\Phi_D$ and $\Phi_U$ transform as a triplet and an antitriplet under 
the global $SU(3)_1$ while ${\cal H}_d$ and ${\cal H}_u$ are a triplet and 
an antitriplet under $SU(3)_2$. The Higgs multiplets should acquire vacuum 
expectation values (vevs) aligned in such a way that the $SU(2)_W$ gauge 
symmetry remains unbroken
\begin{eqnarray}
\langle\Phi_D\rangle=\left(\matrix{0\cr0\cr F_D}\right)~,
\phantom{aaaaaaaaaa} 
\langle{\cal H}_d\rangle=\left(\matrix{0\cr0\cr f\cos\beta}\right)~,
\end{eqnarray}
\begin{eqnarray}
\langle\Phi_U\rangle=\left(0,0,F_U\right)~,
\phantom{aaaaaaaaaa}
\langle{\cal H}_u\rangle=\left(0,0,f\sin\beta\right)~.
\nonumber
\end{eqnarray}

For $F \equiv \sqrt{(F_U^2+F_D^2)/2}\gg f$ we get then the following picture. The 
$SU(3)_W$ gauge symmetry and the global $SU(3)_1$ symmetry are broken down to 
$SU(2)$ at the scale $F$, while the global $SU(3)_2$ survives 
and is spontaneously broken only at the scale $f$. This pattern 
of gauge and global symmetry breaking leads to ten Goldstone bosons, five of 
which become longitudinal components of the massive gauge bosons 
corresponding to broken $SU(3)_W$ generators. For $F\gg f$ the five physical 
Goldstone bosons are dominantly linear combinations of the components of 
${\cal H}_u$ and ${\cal H}_d$. They can be conveniently parametrized as 
follows:   
\begin{eqnarray}
{\cal H}_u=f\sin\beta\left({H^\dagger\over|H|}\sin\left({|H|\over f}\right),~ 
e^{i\eta\over f\sqrt2}\cos\left({|H|\over f}\right)\right)~,
\phantom{aa}
{\cal H}_d=f\cos\beta\left(\matrix{{H\over|H|}\sin\left({|H|\over f}\right)\cr 
e^{-{i\eta\over f\sqrt2}}\cos\left({|H|\over f}\right)}\right)~.
\label{eqn:parametrization}
\end{eqnarray}
Here, as in \cite{CHFAPOWA}, $H$ is a weak $SU(2)_W$ doublet, which is 
identified with the SM Higgs doublet, and $|H|=\sqrt{H^\dagger H}$. 
The remaining Goldstone boson $\eta$ is a SM singlet. We ignore it in most 
of the following discussion, yet we will comment on its physical effects at the 
end of the paper.  

As we outlined in the introduction, we are interested in a scenario in which  
the global symmetry breaking scale $f$ is $f\sim1$~TeV. Obviously, 
supersymmetry should not be broken spontaneously at the scale $F$. 
The required pattern can be obtained by choosing the following superpotential 
for the Higgs sector \cite{FAY}  
\begin{eqnarray}
W = \kappa_1 N_1\left(\Phi_U\Phi_D-\mu^2\right)
  + \kappa_2 N_2{\cal H}_u{\cal H}_d + {1 \over 3} \lambda_2 N_2^3~.\label{eqn:superpot}
\end{eqnarray}
where we introduced singlet superfields $N_{1,2}$ (superfields, and their 
scalar components are denoted by the same letters) and $\mu$ is a mass
parameter. Note that the terms $\Phi_U{\cal H}_d$ and/or ${\cal H}_u\Phi_D$ 
are not present by construction as they  would break the global 
$SU(3)_1\times SU(3)_2$ symmetry. Of course, in supersymmetric models the 
form of the superpotential is stable with respect to radiative corrections 
due to the non-renormalization theorem, even if it is not the most general 
one consistent with the gauge symmetry. Its constrained form could be, in 
principle, a consequence of the local symmetry structure of the high energy 
completion or of some discrete symmetries, but in this paper we do not 
construct any explicit models which would ensure this.

The superpotential (\ref{eqn:superpot}) is a simple choice which has the  
necessary qualitative features. In the limit of unbroken supersymmetry the 
scalar potential resulting from (\ref{eqn:superpot}) has its minimum for 
\begin{eqnarray}
F_U = F_D = \mu 
\end{eqnarray}
and vanishing vacuum expectation values of the other fields.
When the soft masses are taken into account,
\begin{equation}
\label{e.fud}
{\cal L}_{\rm soft} = - M_U^2 |\Phi_U|^2 - M_D^2 |\Phi_D|^2 - M_{N}^2 |N_1|^2 
 - m_u^2 |{\cal H}_u|^2 - m_d^2 |{\cal H}_d|^2 - m_{n}^2 |N_2|^2 \, , 
\end{equation}
the vevs $F_U$, $F_D$ are shifted by terms of order $M_{\rm soft}^2/F$, and $F_U^2 - F_D^2 \sim M_U^2 - M_D^2$.
In the limit $F \gg f \sim M_{\rm soft}$, in order to study the dynamics of light fields  
 we can first decouple the heavy components of $\Phi_U$, $\Phi_D$ and $N_1$ with masses of order $F$.
This procedure yields the effective potential  
\begin{eqnarray} 
\label{e.effpot}
&
V_{\rm eff} = 
|\kappa_2 N|^2 (|{\cal H}_u|^2 + |{\cal H}_d|^2) + |\kappa_2  {\cal H}_u {\cal H}_d + \lambda_2 N_2^2|^2
& \nonumber \\ &
+ {g^2 \over 8}  \left ( {\cal H}_d^\dagger \lambda^i {\cal H}_d - {\cal H}_u \lambda^i {\cal H}_u^\dagger \right)^2 
 + {g_y^2 \over 2} \left ( {\cal H}_d^\dagger \lambda^y {\cal H}_d - {\cal H}_u \lambda^y {\cal H}_u^\dagger \right)^2
& \nonumber \\ &
 + m_u^2 |{\cal H}_u|^2 + m_d^2 |{\cal H}_d|^2 + m_{n}^2 |N_2|^2 + \delta V_{\rm soft} + {\cal O}(1/F^2)
\end{eqnarray}
\begin{equation} 
\label{e.su3b}
\delta V_{\rm soft} \sim (M_D^2 - M_U^2) 
\left[ v_1 ({\cal H}_d^\dagger \lambda^8 {\cal H}_d - {\cal H}_u \lambda^8 {\cal H}_u^\dagger)  
+ v_2 ( {\cal H}_d^\dagger \lambda^y {\cal H}_d - {\cal H}_u \lambda^y {\cal H}_u^\dagger)  \right ]
\end{equation}
Here 
$\lambda^y = {\rm diag}(1/2,1/2,0)$, $\lambda^a$ denote  the Gell-Mann matrices, $i = 1 \dots 3$, 
$g$ and $g_x$ are the gauge couplings of $SU(3)_W \times U(1)_X$, $g_y = {g g_x \over \sqrt{g^2 + g_x^2/3}}$ is the hypercharge coupling and $v_1$, $v_2$ are irrelevant numerical factors.  
We have neglected possible trilinear soft terms. 
In general, the effective potential contains the soft masses (\ref{e.su3b}) that do not respect the global  $SU(3)_2$ \cite{DIPO}, and would give a large tree-level mass to the Higgs doublet.  
To avoid this, we require $M_U^2 \approx M_D^2$  at the scale $F$. 
This is possible, for example, in models with universal soft masses at the supersymmetry breaking mediation scale, 
as long as non-universal contribution  from renormalization group running down to $F$ are small enough. 
The second line in (\ref{e.effpot}) is  the D-term potential corresponding to the unbroken gauge group  
$SU(2)_W \times U(1)_Y$ (the D-terms corresponding to the broken generators of $SU(3)_W \times U(1)_X$ cancel out, when the heavy fields are properly integrated out). 
For $M_U^2 = M_D^2$ these D-terms are the only $SU(3)_2$ breaking terms in the tree-level effective Higgs potential below $F$. Thus, at tree-level we get only the quartic term (and no soft mass term) for the Higgs doublet.  

Soft masses  may induce vevs of the the electroweak singlet components of  ${\cal H}$, so that $f \sim M_{\rm soft}$.
For large $\tan\beta$ we find
\begin{eqnarray}
f^2\approx-{m_n^2\over\kappa_2^2}~,\phantom{aaaa}
\langle N_2\rangle^2\approx-{m_u^2\over\kappa_2^2}~,\phantom{aaaa} 
\tan\beta\approx{\kappa_2\over\lambda_2}{m_n^2+m_u^2-m_d^2\over-m_u^2}~,
\end{eqnarray}
where $m_n$, $m_u$ and $m_d$ are soft masses of $N_2$, ${\cal H}_u$ and 
${\cal H}_d$, respectively. For generic soft masses, large enough $\tan\beta$ 
is obtained for $|\kappa_2|\gg|\lambda_2|$. The necessary negative masses 
squared may result from renormalization group running. Indeed, similarly as 
in the MSSM, the triplet ${\cal H}_u$ mass is driven negative by the large 
top Yukawa coupling, while the singlet soft mass also acquires large 
negative contribution from Yukawa interactions as long as $|\kappa_2|\sim1$.

The $SU(3)_1\times SU(3)_2$ symmetry must only be approximate so that the 
Higgs doublet $H$ is rather a pseudo-Goldstone boson. As we mentioned, the gauge 
interactions below the scale $F$ do not respect the global symmetry, and the 
corresponding $D$-terms generate the quartic Higgs potential. We require 
that another source of the explicit breaking comes from the supersymmetric 
interactions in the top sector. In such case, the Higgs doublet can acquire 
negative mass parameter at one-loop level. However due to an interplay 
between the approximate global symmetry and supersymmetry these radiative 
corrections are finite at one loop \cite{CHFAPOWA}. Logarithmic divergences 
are cut-off by the contribution of the additional top quarks, whose presence 
is required by the approximate $SU(3)$. This double protection mechanism 
alleviates the supersymmetric fine-tuning problem, as we demonstrate in the 
following.

We shall illustrate our point in a specific model, which is a straightforward 
supersymmetrization of "the simplest little Higgs" model of ref. \cite{SCHM}  
and later comment on the generality of our results. The relevant for us chiral 
fermion superfields are the $SU(3)_W$ triplet $\Psi_Q = (Q,T)^T$, and the 
$SU(3)_W$ quark singlets $t^c$ and $T^c$. The superpotential is given by 
\begin{eqnarray}
W = y_1\Phi_U\Psi_Q T^c + y_2{\cal H}_u\Psi_Q t^c~.
\label{eqn:tops}
\end{eqnarray}
As for the Higgs fields, this is not the most general choice consistent with 
the gauge symmetry. Once the $SU(3)_W$ gauge symmetry is broken at the scale 
$F$, the second term in eq.~(\ref{eqn:tops}) preserves the global $SU(3)_2$ 
symmetry, while the first term breaks it explicitly.

Inserting the parametrization (\ref{eqn:parametrization}) of 
${\cal H}_{u,d}$ we can read the top matrix ${\cal L}=-(t,T)
{\cal M}_{\rm top}(t^c,T^c)$ as a function of the vev of $H$. 
The mass matrix  squared reads  
\begin{eqnarray}
{\cal M}_{\rm top}^\dagger{\cal M}_{\rm top}= 
\left(\matrix{y_1^2F^2 & y_1y_2Ff\sin\beta\cos(|H|/f)\cr 
y_1y_2Ff\sin\beta\cos(|H|/f) & y_2^2f^2\sin^2\beta}\right)~.
\label{eqn:m2tops}
\end{eqnarray}
For $|H|\ll f$ the matrix (\ref{eqn:m2tops}) has two hierarchical eigenvalues 
corresponding to the standard model top quark and its heavy $SU(3)_W$ partner:
\begin{eqnarray}
m_t&\approx&y_t|H|~, \phantom{aaaaa} y_t = {y_1y_2F\over m_T}\sin\beta~, 
\nonumber\\
m_T&\approx&\sqrt{y_1^2 F^2 + y_2^2 f^2\sin^2\beta}~.
\label{eqn:topmasses}
\end{eqnarray}

At this point it is useful to summarize the orders of magnitudes of the model 
parameters needed for a coherent picture. The scale $F$  cannot be too high 
because soft terms must be approximately $SU(3)$ symmetric at the TeV scale. 
For definiteness we will assume $F\sim10$~TeV. On the other hand, for the 
sake of minimizing the fine-tuning we will need $m_T\sim1$~TeV, so that we 
need $y_1F\sim y_2 f\simlt1$~TeV. 
Note that the smallness of $y_1$ is consistent with our assumption that renormalization group effects do not generate a large splitting between $M_U^2$ and $M_D^2$.
 For given values of $F$ and $f$ the Yukawa 
couplings $y_1$ and $y_2$ can be chosen such that the Standard Model top 
Yukawa coupling $y_t$ has the correct value and $m_T$ is in the desired range. 
Relation (\ref{eqn:topmasses}) implies then a lower limit $m_T>2y_tf$.  

We move to the determination of the Higgs potential
\begin{eqnarray}
V=\delta m_H^2 |H|^2 + (\lambda_0+\delta\lambda)|H|^4 + \dots 
\end{eqnarray}
The tree-level quartic term comes from $SU(2)_W\times U(1)_Y$ 
$D$-terms.
Its form is analogous as in the MSSM and, in the limit $\tan\beta\gg1$, reads
\begin{eqnarray}
\lambda_0 = {g^2 + g_y^2\over8}~.
\end{eqnarray}
The one-loop corrections $\delta m_H^2$ and $\delta\lambda$ can be computed  
{} from the 1-loop effective potential:
\begin{eqnarray}
\Delta V_{\rm1-loop}={1\over64\pi^2}{\rm STr}\left\{{\cal M}^4\left( 
\ln{{\cal M}^2\over\Lambda_{UV}^2}-{3\over2}\right)\right\}~.
\end{eqnarray}
By computing ${\rm STr}{\cal M}^4$ it can be seen that there is no 
logarithmically divergent contribution from the top and stop sector to the 
Higgs potential mass parameter $\delta m^2_H$. This can be understood by a 
simple dimensional analysis. The coefficient of a logarithmically divergent 
term would have to break both supersymmetry and the approximate global 
symmetry, but in the top-stop sector there is no such dimensionful parameter. 
The conclusion holds for any stop soft masses and trilinear terms, as long as 
their tree-level values respect the $SU(3)_2$ global symmetry. Furthermore, 
in a simplified situation when all stops have approximately the same soft 
mass squared $m_Q$ and the mixing between  left and right-chiral stops is 
negligible we obtain a simple formula for the Higgs mass parameter:  
\begin{eqnarray}
\delta m_H^2\approx-{3\over8\pi^2}y_t^2\left[(m_Q^2 + m_T^2)
\ln(m_Q^2 + m_T^2) - m_Q^2\ln m_Q^2 - m_T^2\ln m_T^2\right] + \Delta~.
\label{eqn:mH2}
\end{eqnarray}
Here $\Delta$ stands for contributions from other sectors of the theory. 
For example, the SM gauge interactions contribute   
\begin{eqnarray}
\Delta\supset{3g_2^2M_2^2+g_y^2M_y^2\over8\pi^2}\ln{F\over M_{\rm soft}}~,
\end{eqnarray}
where $M_2$ and $M_y$ are soft gaugino masses. The cut-off is given by the 
scale at which the $SU(3)_W$ gauge symmetry is restored.

The top contribution in (\ref{eqn:mH2}) has a remarkable property that it 
vanishes for both $m_Q\rightarrow0$ and $m_T\rightarrow0$. As advertised,
the double protection mechanism leads to the softening of the UV sensitivity 
of the Higgs potential. 
For a given $m_T$, the top contribution is 
minimized for $m_Q=m_T$, while for $m_Q>m_T$ it increases only as 
$\ln(m_Q/m_T)$. 

The dominant contribution to the Higgs potential quartic coupling (needed 
to evaluate the Higgs boson mass) is given by
\begin{eqnarray}
\delta\lambda\approx{3\over16\pi^2}y_t^4\left[
\ln\left({m_T^2 m_Q^2\over(m_Q^2 + m_T^2)m_t^2}\right)+{3\over2} 
-2{m_Q^2\over m_T^2}\ln\left({m_Q^2 + m_T^2\over m_Q^2}\right)\right]~.
\label{eqn:dlambda}
\end{eqnarray}
For $m_Q\sim m_T$ it behaves as $(3/8\pi^2)y_t^4\ln(m_Q/m_t)$, very much 
like in the MSSM. Therefore, the physical Higgs boson mass given by 
\begin{eqnarray}
M_h^2 = 2(\lambda_0 + \delta\lambda)v^2 = M_Z^2 + 2\delta\lambda v^2
\label{eqn:higgsmass}
\end{eqnarray}
takes similar values as in the MSSM with analogous stop masses. 

We are now ready to estimate the level of fine-tuning of the electroweak 
breaking. The value of $\delta m_H^2$ is tied to the electroweak scale by 
the relation $v^2 = -\delta m_H^2/(\lambda_0+\delta\lambda)$, where 
$v=246$~GeV. We can always obtain the correct $v$ by arranging for appropriate 
$\Delta$ (for example, by tuning the gaugino masses), but large cancellations 
between the top contribution and $\Delta$ are unnatural.   
We can quantify the fine-tuning  as follows:
\begin{eqnarray}
{\rm FT}=\left|{\Delta - |\delta m_H^2|\over|\delta m_H^2|}\right| 
= \left|{\Delta-{1\over 2}M_h^2\over {1\over 2}M_h^2}\right|
\end{eqnarray}

\begin{figure}[t]
\begin{center}
\centerline{\includegraphics[width=0.45\textwidth]{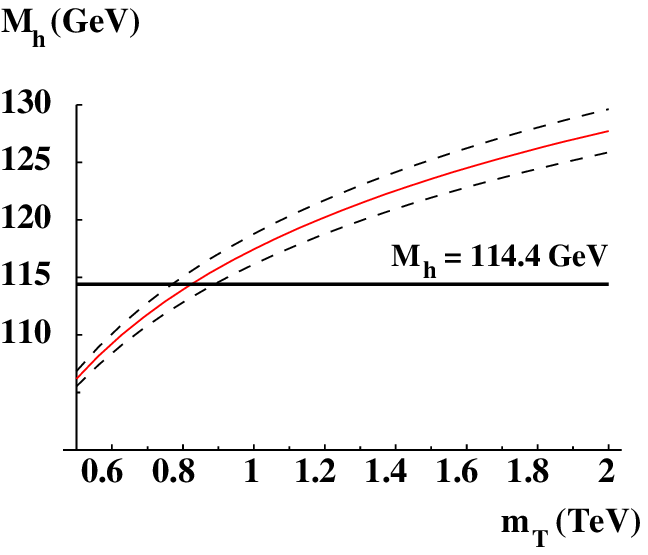}
            \includegraphics[width=0.45\textwidth]{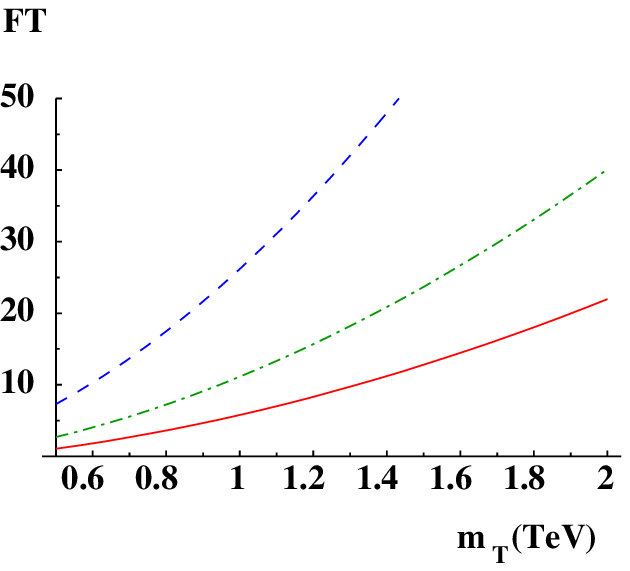}}
\caption{{\footnotesize 
Left panel: The Higgs boson mass  as a function of $m_T$ for $m_Q/m_T=1$. 
The dashed lines  indicate the effect of the $2\sigma$ 
uncertainty of the top mass.
Right panel:  the fine-tuning as a function of $m_T$ for $m_Q/m_T=1$ (solid red), $m_Q/m_T=2$ (dot-dashed green),  $m_Q/m_T=10$ (dashed blue). }}
\label{fig:ft}
\end{center}
\end{figure}

In fig.~\ref{fig:ft} we show the dependence of the Higgs boson mass and the 
fine-tuning on the input parameters of our model. The Higgs boson mass is 
plotted as a function of $m_T$ for $m_Q=m_T$ (the dependence on the $m_Q/m_T$ 
ratio is very weak in this case). This is compared with the direct search 
limit from LEP2, $M_h>114.4$~GeV. We have used the value of the 
top mass $m_t = 172.7\pm2.9$~GeV \cite{FERMILAB} and the corresponding 
$\overline{\rm MS}$ central value $164\pm3$~GeV. The effect of varying the 
top mass within the $2\sigma$ limit is also displayed. The fine-tuning as a 
function of $m_T$ is plotted for several values of the ratio $m_Q/m_T\geq1$ 
(the formulae are of course symmetric under interchange of $m_Q$ and $m_T$). 
We conclude that for 0.8~TeV$<m_T<1$~TeV and $m_Q\sim m_T$ the electroweak 
symmetry can be broken with no fine-tuning at all, while the LEP2 Higgs mass 
bound can be respected. For $m_T> 1$~TeV the fine-tuning is of order $10\%$. 
Note that even when either $m_Q$ or $m_T$ are of oder $10$~TeV fine-tuning 
is not worse than in the MSSM with TeV scale stop masses.  

We leave the detailed analysis of phenomenological properties of the double 
protection mechanism for a future publication. Here we just point out its 
main experimental signatures. One of them is the existence of a vector-like 
top quark with mass around 1~TeV. New gauge bosons are expected to be much 
heavier, not within the reach of the LHC. The gauginos should have masses at 
most around 1~TeV. Furthermore, if the mechanism of spontaneous global 
symmetry breaking at the scale $f$ is linked to the soft supersymmetry 
breaking parameters, as in the mechanism discussed in this paper, squark 
masses are around 1~TeV, too. However, on can perhaps think about other 
mechanisms of generating the scale $f$. It is worth noting that merely from 
the point of view of the electroweak symmetry breaking with little fine 
tuning of the parameters, the squark masses are very weakly constrained. 

We now comment on the pseudoscalar singlet $\eta$ in the parametrization 
(\ref{eqn:parametrization}). It is a true massless Goldstone boson 
corresponding to the Peccei-Quinn $U(1)$ symmetry acting on the $SU(3)$ 
partners of the SM weak doublets. As such, it is subject to experimental, 
cosmological and astrophysical constraints on light bosons \cite{PDGax}. 
However, $\eta$ couples to the ordinary matter only via mixing of the 
latter with their $SU(3)$ partners. For the first two generation such 
mixing can be very small, as long as the corresponding $SU(3)$ partners 
have masses of order $f$. Therefore, processes like energy loss in stars 
\cite{PDGax} put on $f$ only a weak lower bound of order 100~GeV.
At the nucleosynthesis epoch, the background $\eta$'s, decoupled from the 
thermal bath would contribute to the energy density of the Universe. The 
cosmological effect of such a scalar is equivalent to the one of 0.57 
neutrino generation. The conservative estimates of $N_\nu$ from 
nucleosynthesis still allow for $1.4<N_\nu <4.9$ \cite{PDGbbn}, and do not 
exclude the presence of $\eta$.

The model studied above can be extended so as to accommodate all three 
generations of quark and leptons and their masses \cite{SCHM}. In this case, 
anomaly cancellation implies that the assignment to $SU(3)_W \times U(1)_X$ 
representations cannot be generation universal. Furthermore, it is not 
possible to embed this spectrum in a simple unified group. However one 
can consider models with a different spectrum, which ensure the double 
protection mechanism for the Higgs potential. For example, with the 
$SU(3)_C\times SU(3)_W\times U(1)_X$ gauge symmetry the following matter 
content can be chosen (the first two generations can be introduced 
analogously):
\begin{eqnarray}
\Phi_U,~{\cal H}_u: ~(1,3)_{1/3} \phantom{aaaa} 
\Phi_D,~{\cal H}_d: ~(1,\bar3)_{-1/3}
\phantom{aaaaaaaaaa}\nonumber\\
\psi_Q=\left(\matrix{Q_3\cr D}\right): ~(3,3)_0
\phantom{aaaa}
T^c: ~(\bar3,3)_{-1/3} 
\phantom{aaaa}
T: ~(3,\bar3)_{1/3} 
\phantom{aaaa}
t^c: ~(\bar 3,1)_{-2/3}\\
b_1^c,~b_2^c:~(\bar 3,1)_{1/3}
\phantom{aaaa}
\psi_L =\left(\matrix{L_3\cr E}\right): ~(1,3)_{-2/3}
\phantom{aaaa}
\tau_1^c,~\tau_2^c: ~(1,1)_1\nonumber
\end{eqnarray} 
with the superpotential 
\begin{eqnarray}
W &=& y_1 \Phi_U \Psi_Q T^c + y_2 {\cal H}_u T t^c + \mu_T T T^c
\nonumber\\
&+&y_{b1} \psi_Q {\cal H}_d b_1^c +  y_{b2} \psi_Q \Phi_D b_2^c 
+ y_{\tau 1} \psi_L {\cal H}_d \tau_1^c 
+  y_{\tau 2} \psi_L \Phi_D \tau_2^c \, .
\end{eqnarray}
The top sector here is slightly more complicated. It contains a vector like 
triplet pair $T$ and $T^c$ with a supersymmetric mass term $\mu_T$. Still, 
for $F\gg f$ the picture is qualitatively and quantitatively the same as in 
the supersymmetric version of the model of ref. \cite{SCHM} discussed in 
this paper. In particular, the 
formulae (\ref{eqn:mH2}) and (\ref{eqn:dlambda}) for the parameters of the 
Higgs potential still hold, with $y_1 F$ replaced by $\mu_T$ in 
eq.~(\ref{eqn:topmasses}). This indicates that the structure of the effective 
Higgs potential at one loop is a general feature of models in which
the double protection mechanism is realized.  

In conclusion, the double protection of the Higgs potential: by supersymmetry 
softly broken at the TeV scale
and by a global symmetry which is spontaneously broken at the scale 
$\simlt1$~TeV may be a mechanism allowing to understand the origin and the 
stability of the Fermi scale.

\vskip1.0cm

{\bf Note added:} 
Shortly after our paper appeared, models with double protection of the Higgs potential were discussed by Roy and Schmaltz in ref. \cite{SZMALC}. 
These authors consider a similar model like the one in our paper and conclude it is not viable. 
In the model  of ref. \cite{SZMALC} the conditions $M_U^2 = M_D^2$ and $m_u^2 = m_d^2$ are imposed
and the scale $f$ is generated by supersymmetric terms in the potential (and not by soft terms, as in our model). 
Under these assumptions 
$F_U = F_D$ and  $\tan \beta = 1$, 
which ensure that the D-term contributions to the Higgs mass parameter are absent.
Obviously,  for $\tan \beta  = 1$ the Higgs boson is not heavy enough.  
However, it has been overlooked in \cite{SZMALC}  that for $F \gg f$ the condition $\tan \beta = 1$ is not needed 
 to avoid  the D-term contributions to the Higgs mass parameter. 
This is discussed  in our paper below eq. (\ref{e.fud}). 

\vskip1.0cm

\section*{Acknowledgments}

P.H.Ch. thanks K.A. Meissner for a discussion and the CERN Theory for 
hospitality during the completion of this paper.
Z.B. was partially supported by the MIUR grant for the Projects of National
Interest PRIN 2004 for ``Astroparticle Physics''. 
P.H.Ch., A.F. and S.P. were partially supported by the European Community 
Contract MRTN-CT-2004-503369 for years 2004-2008 and by the Polish KBN 
grant 1 P03B 099 29 for years 2005--2007. The stay of A.F. at DESY is 
possible owing to the Research
Fellowship granted by the Alexander von Humboldt Foundation.



\begin{thebibliography}{99}


\bibitem{LITHIG} N.~Arkani-Hamed, A.~G.~Cohen and H.~Georgi,
Phys.\ Lett.\ B {\bf 513} (2001) 232 [arXiv:hep-ph/0105239];\\
N.~Arkani-Hamed, A.~G.~Cohen, E.~Katz, A.~E.~Nelson, T.~Gregoire and 
J.~G.~Wacker, JHEP {\bf 0208} (2002) 021 [arXiv:hep-ph/0206020];\\
N.~Arkani-Hamed, A.~G.~Cohen, E.~Katz and A.~E.~Nelson,
JHEP {\bf 0207} (2002) 034 [arXiv:hep-ph/0206021];
M.~Schmaltz and D.~Tucker-Smith,
arXiv:hep-ph/0502182.

\bibitem{CASAS} J.~A.~Casas, J.~R.~Espinosa and I.~Hidalgo, 
JHEP {\bf 0503} (2005) 038 [arXiv:hep-ph/0502066].

\bibitem{STSCH} G.~Marandella, C.~Schappacher and A.~Strumia,  
Phys.\ Rev.\ D {\bf 72} (2005) 035014 [arXiv:hep-ph/0502096].

\bibitem{CHFAPOWA} P.~H.~Chankowski, A.~Falkowski, S.~Pokorski and J.~Wagner,
Phys.\ Lett.\ B {\bf 598} (2004) 252 [arXiv:hep-ph/0407242].

\bibitem{BICHGA} A.~Birkedal, Z.~Chacko and M.~K.~Gaillard, JHEP {\bf 0410}, 
036 (2004) [arXiv:hep-ph/0404197].
  
\bibitem{SCHM} M.~Schmaltz, JHEP {\bf 0408}, 056 (2004) [arXiv:hep-ph/0407143].

\bibitem{FRAMP}
P.~H.~Frampton,
Phys.\ Rev.\ Lett.\  {\bf 69} (1992) 2889;
F.~Pisano and V.~Pleitez,
Phys.\ Rev.\ D {\bf 46} (1992) 410
[arXiv:hep-ph/9206242];
D.~E.~Kaplan and M.~Schmaltz,
JHEP {\bf 0310} (2003) 039
[arXiv:hep-ph/0302049];
W.~Skiba and J.~Terning,
Phys.\ Rev.\ D {\bf 68} (2003) 075001
[arXiv:hep-ph/0305302];
O.~C.~W.~Kong,
arXiv:hep-ph/0307250.


\bibitem{FAY} 
P.~Fayet,
Nucl.\ Phys.\ B {\bf 90} (1975) 104.
 
\bibitem{DIPO}
  A.~Pomarol and S.~Dimopoulos,
  Nucl.\ Phys.\ B {\bf 453} (1995) 83
  [arXiv:hep-ph/9505302];
  R.~Rattazzi,
  Phys.\ Lett.\ B {\bf 375} (1996) 181
  [arXiv:hep-ph/9507315].

\bibitem{FERMILAB} The CDF and D0 Collaborations and the Tevatron 
Electroweak Working Group arXiv:hep-ex/0507091.

\bibitem{PDGax}
G. Raffelt, in Review of Particle Physics, S.~Eidelman {\it et al.}  
[Particle Data Group] Phys.\ Lett.\ B {\bf 592} (2004) 1.

\bibitem{PDGbbn}
B.D.~Fields and S.~Sarkar, in Review of Particle Physics, S.~Eidelman 
{\it et al.}  [Particle Data Group] Phys.\ Lett.\ B {\bf 592} (2004) 1.

\bibitem{SZMALC}
  T.~Roy and M.~Schmaltz,
  arXiv:hep-ph/0509357.

  
    
\end{thebibliography}
\end{document}